\documentclass[prd,twocolumn,showpacs,floatfix,superscriptaddress]{revtex4}

\usepackage{graphicx}
\usepackage{epsfig}
\usepackage{bm}
\usepackage{amssymb}
\usepackage{float}
\usepackage{amsmath}
\usepackage{dcolumn}
\usepackage{cancel}
\usepackage[colorlinks]{hyperref}
\usepackage[usenames,dvipsnames]{color}
\hypersetup{
     breaklinks=true,
    pdfstartview={FitH},    % fits the width of the page to the window
    colorlinks=true,       % false: boxed links; true: colored links
    linkcolor=blue,          % color of internal links
    citecolor=red,        % color of links to bibliography
    filecolor=magenta,      % color of file links
    urlcolor=blue,           % color of external links
    anchorcolor=green,      % Color for anchor text
    linktocpage=true
}

\renewcommand{\)}{\right)}

\def\doi{http://doi.org}
\def\arxiv{http://arxiv.org/abs}

\renewcommand{\)}{\right)}

\renewcommand{\]}{\right]}

\begin{document}

\title{Holographic dark energy through Tsallis entropy}

\author{Emmanuel N. Saridakis}
\email{Emmanuel\_Saridakis@baylor.edu}
\affiliation{Chongqing University of Posts \& Telecommunications, Chongqing, 400065, 
China}
\affiliation{Department of Physics, National Technical University of Athens, Zografou
Campus GR 157 73, Athens, Greece}
\affiliation{ Eurasian  International Center for Theoretical
Physics, Eurasian National University, Astana 010008, Kazakhstan}

\author{Kazuharu Bamba}
\email{bamba@sss.fukushima-u.ac.jp}
\affiliation{Division of Human Support System, Faculty of Symbiotic Systems Science,
Fukushima University, Fukushima 960-1296, Japan}

\author{R. Myrzakulov}\email{rmyrzakulov@gmail.com}
\affiliation{ Eurasian  International Center for Theoretical
Physics, Eurasian National University, Astana 010008, Kazakhstan}

\author{Fotios K. Anagnostopoulos}
\email{fotis-anagnostopoulos@hotmail.com}
\affiliation{National and Kapodistrian University of Athens, Physics Department,
Panepistimioupoli Zografou,  15772, Athens, Greece}
 
\begin{abstract} 
In order to apply holography and entropy relations to the whole universe, which is a 
gravitational and thus nonextensive system, for consistency one should use the 
generalized definition for the universe horizon entropy, namely Tsallis nonextensive 
entropy. We formulate Tsallis holographic dark energy, which is a generalization of 
standard holographic dark energy quantified by a new dimensionless parameter $\delta$, 
possessing the latter as a particular sub-case.  We provide a simple  differential 
equation  for the  dark energy density parameter, as well as an analytical expression for 
its  equation-of-state parameter. In this scenario  the universe exhibits the usual 
thermal history, namely the successive sequence of matter and dark-energy epochs,
before resulting in a complete dark energy domination in the far future. Additionally, 
the 
dark energy equation-of-state parameter presents a rich behavior and,
according to the value of $\delta$, it can be quintessence-like, phantom-like, or 
experience the phantom-divide crossing before or after the present time.  Finally, we 
confront the scenario with Supernovae type Ia and Hubble parameter observational data, 
and we show that the agreement is very good, with $\delta$  preferring a value 
slightly larger than its standard value 1.
\end{abstract}

\pacs{98.80.-k, 95.36.+x, 04.50.Kd}

\maketitle

\section{Introduction}

According to the concordance paradigm of cosmology the universe experienced an early-time 
accelerated phase, followed by the sequence of radiation and matter dominated eras, 
before resulting in the current, late-time, accelerated epoch. The two accelerated phases 
cannot be easily described through general relativity and standard model of particle 
physics, since extra degree(s) of freedom seem to be necessarily required. From one hand 
we can attribute these extra degrees of freedom to new, exotic forms of matter, 
collectively named as dark energy  \cite{Copeland:2006wr,Cai:2009zp,Bamba:2012cp}. On the 
other 
hand we can consider them to be of gravitational origin, namely to arise from a modified 
theory of gravity that includes  general relativity as a low-energy limit  
\cite{Nojiri:2006ri,Nojiri:2010wj,Capozziello:2011et,Cai:2015emx,Nojiri:2017ncd}.

One interesting alternative for the explanation of dark energy origin and nature can be 
acquired applying the holographic principle 
\cite{tHooft:1993dmi,Susskind:1994vu,Bousso:2002ju} at a cosmological framework 
\cite{Fischler:1998st,Bak:1999hd,Horava:2000tb}. In particular, one takes advantage of 
the 
connection between the  Ultraviolet cutoff of the (quantum field) theory, which is 
related 
to the vacuum energy, with the (necessary for the applicability of the quantum field 
theory at large distances) largest distance of the theory \cite{Cohen:1998zx}. In 
this way the resulted vacuum energy will be a form of dark energy of holographic origin, 
named holographic dark energy \cite{Li:2004rb} (see \cite{Wang:2016och} for a review). 
Holographic dark energy leads to interesting cosmological phenomenology  
\cite{Li:2004rb,Wang:2016och,Horvat:2004vn,Huang:2004ai,Pavon:2005yx,Wang:2005jx,
Nojiri:2005pu,Kim:2005at,
Wang:2005ph, Setare:2006wh,Setare:2008pc,Setare:2008hm} and it has been also extended 
through various ways 
\cite{Gong:2004fq,Saridakis:2007cy,  
Setare:2007we,Cai:2007us,Setare:2008bb,Saridakis:2007ns,Saridakis:2007wx,Jamil:2009sq,
Gong:2009dc, 
Suwa:2009gm,Jamil:2010vr,BouhmadiLopez:2011xi,Malekjani:2012bw,Khurshudyan:2014axa,
Landim:2015hqa,Pasqua:2015bfz,
Jawad:2016tne,Pourhassan:2017cba,Nojiri:2017opc,Saridakis:2017rdo}. Additionally, 
holographic dark energy 
can be shown to be in agreement  with   
observational data 
\cite{Zhang:2005hs,Li:2009bn,Feng:2007wn,Zhang:2009un,Lu:2009iv,Micheletti:2009jy}.

A crucial ingredient of the cosmological application of holography is the fact that 
the entropy  of the whole universe, considered as a system with 
radius the aforementioned largest distance, is proportional to its area, similarly to 
a black hole. However, already at 1902 Gibbs had pointed out that in systems in which 
the partition function diverges the Boltzmann-Gibbs theory cannot be applied, and we now 
know that gravitational systems lie within this class. As it was shown by Tsallis, in 
such cases the usual Boltzmann-Gibbs additive entropy (which is  
 founded on the hypothesis of weak probabilistic correlations and their connections to 
ergodicity) must be generalized to the non-additive entropy (i.e the entropy of the 
whole system is not necessarily the sum of the entropies of its sub-systems), know as 
Tsallis 
entropy  \cite{Tsallis:1987eu,Lyra:1998wz,Tsallis:1998ws,Wilk:1999dr}. In particular, 
this 
 nonextensive Tsallis entropy can be written in compact form as \cite{Tsallis:2012js}
\begin{equation}
\label{Tsallisentropy}
S_T=\gamma A^\delta,
\end{equation}
where $A\propto L^2$ is the area of the system with characteristic length $L$. The 
parameters $\gamma$ and $\delta$ under the hypothesis of equal 
probabilities are related to the dimensionality of the system $d$, 
and specifically the important one is $\delta=d/(d-1)$ for $d>1$ 
\cite{Tsallis:2012js}, 
however in the general case they remain as completely free parameters. Obviously, in the 
case where $\delta=1$ and $\gamma=2 \pi M_p^2$ (in units where $\hbar=k_B = c = 1$), with 
$M_p$ the Planck mass, we obtain the usual additive entropy.

Having these in mind, we deduce that in order to apply holography and entropy relations 
to the whole universe, which is a gravitational and thus nonextensive system, one should 
use the above generalized definition of the universe horizon entropy. Hence, in the case 
of holographic dark energy, which is obtained from the inequality $\rho_{DE} 
L^4\leq S$ with $S\propto A\propto L^2 $ \cite{Wang:2016och}, the consistent scenario 
will 
arise if we use the Tsallis entropy (\ref{Tsallisentropy}) in this inequality, resulting 
to
 \begin{equation}
\label{THDE}
\rho_{DE}=B L^{2\delta-4},
\end{equation}
with $B$ a parameter with dimensions  $[L]^{-2\delta}$. As mentioned above, for 
$\delta=1$ the above expression gives the 
usual holographic dark energy  $\rho_{DE}=3c^2 M_p^2 L^{-2}$, with $B=3 c^2 M_p^2$ and 
$c^2$ the model parameter. Additionally, it is worth mentioning that in the special case 
$\delta=2$ the above relation gives the standard cosmological constant case 
$\rho_{DE}=const.=\Lambda$.

In the present work we are interested in formulating Tsallis holographic dark energy, 
which is characterized by energy density (\ref{THDE}), and investigate its cosmological 
implications. Although relation (\ref{THDE}) has been also extracted in a recent work too 
\cite{Tavayef:2018xwx}, its cosmological application has the serious disadvantage that it 
does not possess standard entropy and standard holographic dark energy as a sub-case. 
The 
reason behind this failure is the fact that it was the Hubble horizon that was used as 
$L$ (see also \cite{Jahromi:2018xxh,Moradpour:2018ivi} where the same inconsistency 
appears), which is well known that cannot lead to realistic cosmology in the case of 
usual 
holographic dark energy \cite{Hsu:2004ri}. Hence, in the present paper we proceed to a 
consistent 
formulation of Tsallis holographic dark energy, taking as $L$ the future event horizon, 
namely the same length that is used in standard  holographic dark energy scenario. In 
this way Tsallis holographic dark energy is indeed a consistent generalization of 
standard 
holographic dark energy, possessing it as a particular limit, namely for $\delta=1$.

The plan of the manuscript is the following. In Section \ref{model} we formulate 
Tsallis holographic dark energy in a consistent way, extracting the corresponding 
cosmological equations. In Section \ref{Cosmologicalevolution} we investigate the 
cosmological  behavior of the scenario, focusing on the evolution of the dark-energy 
density and equation-of-state parameters, and we  confront it with Supernovae type 
Ia observational data. Finally, Section \ref{Conclusions} is 
devoted to the conclusions.

\section{Tsallis holographic dark energy}
\label{model}
  
In this section we present the basic expressions of holographic dark energy based on 
Tsallis nonextensive entropy. Throughout this work we 
consider a flat homogeneous and isotropic Friedmann-Robertson-Walker (FRW) geometry with
  metric  
\begin{equation}
\label{FRWmetric}
ds^{2}=-dt^{2}+a^{2}(t)\delta_{ij}dx^{i}dx^{j}\,,
\end{equation}
with $a(t)$ the scale factor.   

As we mentioned in the Introduction, the starting point for Tsallis holographic dark 
energy is expression (\ref{THDE}). In the formulation of holographic dark energy one 
needs 
to consider a particular IR cutoff, namely the largest length of the theory $L$ that 
appears in the expression of holographic dark energy density. It is well known that in 
the 
case of standard holographic dark energy models $L$ cannot be the Hubble horizon $H^{-1}$ 
(with $H\equiv \dot{a}/a$ the Hubble parameter), since such a choice leads to 
inconsistencies \cite{Hsu:2004ri}. Hence, it was the   future event horizon that was 
finally used \cite{Li:2004rb}, namely 
\begin{equation}
\label{futurehor}
R_h\equiv a\int_t^\infty \frac{dt}{a}= a\int_a^\infty \frac{da}{Ha^2}.
\end{equation}
In a recent attempt to construct Tsallis holographic dark energy the authors used the 
extended relation (\ref{THDE}) but they chose $L$ to be the Hubble horizon 
\cite{Tavayef:2018xwx}. Thus, the resulted model does not have standard holographic 
dark energy and standard thermodynamics as a sub-case, which is a serious disadvantage. 
On the contrary, in the present work we desire to formulate  Tsallis holographic dark 
energy in a consistent way, and hence we use as $L$ the future event horizon 
(\ref{futurehor}). In this way, as we will see, standard holographic dark energy is 
included as a sub-case, and can be obtained for $\delta=1$.

According to the above discussion, and using (\ref{THDE}) with $L$ the $R_h$, the energy 
density of Tsallis holographic dark energy writes as
 \begin{equation}
\label{THDE2}
\rho_{DE}=B R_h^{2\delta-4}.
\end{equation}
In the following we focus on the general case $\delta\neq2$, since as we mentioned for  
$\delta=2$ the model gives the standard cosmological constant $\rho_{DE}=\Lambda$.
The Friedmann equations in a universe containing the dark energy and matter 
perfect fluids are
 \begin{eqnarray}
\label{Fr1b}
3M_p^2 H^2& =& \ \rho_m + \rho_{DE}    \\
\label{Fr2b}
-2 M_p^2\dot{H}& =& \rho_m +p_m+\rho_{DE}+p_{DE},
\end{eqnarray}
with $p_{DE}$ the pressure of  Tsallis holographic dark energy, and $\rho_m$ and $p_m$ 
respectively the energy density and pressure of the matter sector.
The equations close by considering the matter
conservation 
equation  
\begin{equation}\label{rhoconserv}
\dot{\rho}_m+3H(\rho_m+p_m)=0.
\end{equation}

It proves convenient to introduce the dark energy and matter density parameters through
 \begin{eqnarray}
 && \Omega_m\equiv\frac{1}{3M_p^2H^2}\rho_m
 \label{Omm}\\
 &&\Omega_{DE}\equiv\frac{1}{3M_p^2H^2}\rho_{DE}.
  \label{ODE}
 \end{eqnarray}
 Using these definitions, relations (\ref{futurehor}),(\ref{THDE2}) lead to
  \begin{equation}\label{integrrelation}
\int_x^\infty \frac{dx}{Ha}=\frac{1}{a}\left(\frac{B}{3M_p^2H^2\Omega_{DE}}
\right)^{\frac{1}{4-2\delta}},
\end{equation}
 where  $x\equiv \ln a$. 
 In the following we focus on the dust matter case, namely we consider the matter 
equation-of-state parameter  to be zero, and thus 
(\ref{rhoconserv})  
gives $\rho_m=\rho_{m0}/a^3$, with $\rho_{m0}$ the value of the matter energy density  
at the present scale factor $a_0=1$ (from now on the subscript ``0'' marks the  
present value of a quantity).  Therefore, inserting into (\ref{Omm}) gives 
$\Omega_m=\Omega_{m0} H_0^2/(a^3 H^2)$, which, using that the Friedmann equation 
(\ref{Fr1b}) becomes  $\Omega_m+\Omega_{DE}=1$, 
 leads to  
 \begin{equation}\label{Hrel2}
\frac{1}{Ha}=\frac{\sqrt{a(1-\Omega_{DE})}}{H_0\sqrt{\Omega_{m0}}}.
\end{equation}

Inserting (\ref{Hrel2}) into (\ref{integrrelation}) we obtain
  \begin{equation}\label{integrrelation2}
\int_x^\infty \frac{dx}{H_0\sqrt{\Omega_{m0}}}  
\sqrt{a(1-\Omega_{DE})}    =\frac{1}{a}\left(\frac{B}{
3M_p^2H^2\Omega_{DE}}
\right)^{\frac{1}{4-2\delta}}.
\end{equation}
It proves convenient to use  $x=\ln a$ as the 
independent variable, and thus for every quantity $f$ we have $\dot{f}=f' H$, where 
primes denote derivatives with respect to $x$. Thus, differentiating 
(\ref{integrrelation2}) with respect to $x$ we finally acquire
  \begin{eqnarray}\label{Odediffeq}
&&
\!\!\!\!\!\!\!\!\!\!\!
\frac{\Omega_{DE}'}{\Omega_{DE}(1-\Omega_{DE})}=2\delta-1+
Q
(1-\Omega_{DE})^{\frac{1-\delta}{
2(2-\delta) } } \nonumber\\
&&\ \ \ \ \ \ \ \ \ \ \ \ \ \  \ \ \ \ \ \ \ \ \ \  \  \ \ \   \ \ \ 
\cdot(\Omega_{DE})^{\frac{1}{2(2-\delta) } } 
e^{\frac{3(1-\delta)}{2(2-\delta)}x},
\end{eqnarray}
 where
   \begin{equation}\label{Qdef}
Q\equiv 2(2-\delta)\left(\frac{B}{3M_p^2}\right)^{\frac{1}{2(\delta-2)}} 
\left(H_0\sqrt{\Omega_{m0}}\right)^{\frac{1-\delta}{\delta-2}}.
\end{equation}

Equation (\ref{Odediffeq}) is the differential equation that determines the 
evolution of Tsallis holographic dark energy, in a flat universe and for dust matter, as 
a function of $x=\ln a$. 
In the case where $\delta=1$ this equation does not have an explicit $x$-dependence and 
it coincides with the one of usual holographic dark energy  \cite{Li:2004rb},  namely 
$\Omega_{DE}'|_{\delta=1}= 
\Omega_{DE}(1-\Omega_{DE})\left(1+2\sqrt{\frac{3M_p^2\Omega_{DE}}{B}}
\right)
$ 
 (complete coincidence is acquired under the identification  
$B=3 
c^2 M_p^2$), which accepts an 
analytic solution in an implicit form  \cite{Li:2004rb}. Nevertheless, in the case 
where $\delta\neq1$, differential equation (\ref{Odediffeq})
exhibits an explicit $x$-dependence and cannot accept an analytical solution. Hence, in 
the following we will elaborate it numerically.

Let us now determine the other important observable, namely the
Tsallis holographic dark energy equation-of-state parameter 
$w_{DE}\equiv p_{DE}/\rho_{DE}$. Since the 
matter sector is conserved, namely Eq. (\ref{rhoconserv}) holds, the two Friedmann 
equations (\ref{Fr1b}),(\ref{Fr2b}) imply that the dark energy sector is conserved too, 
namely 
\begin{equation}\label{rhodeconserv}
\dot{\rho}_{DE}+3H\rho_{DE}(1+w_{DE})=0.
\end{equation}
Differentiating the basic relation (\ref{THDE2}) we obtain that 
$\dot{\rho}_{DE}=2(\delta-2)B R_h^{2\delta-5} \dot{R}_h$, where $\dot{R}_h$ can be 
straightforwardly found from (\ref{futurehor}) to be 
$\dot{R}_h=H  R_h-1$, and where $R_h$ can be eliminated in terms of $\rho_{DE}$ through 
$ R_h=(\rho_{DE}/B)^{1/(2\delta-4)}$, according to (\ref{THDE2}). Inserting these into 
(\ref{rhodeconserv}) we obtain 
\begin{eqnarray}\label{rhodeconserv2}
&&
\!\!\!\!\!\!\!\!\!\!\!\!\!\!\!\!\!\!\!\!\!\!\!\!\!\!\!\!\!\! 
2(\delta-2)B 
\left(\frac{\rho_{DE}}{B}\right)^{\frac{2\delta-5}{2(\delta-2)}}
 \left[H  
\left(\frac{\rho_{DE}}{B}\right)^{\frac{1}{2(\delta-2)}}-1\right]\nonumber\\
&&\ \ \ \ \ \  \ \  \ \ \ \ \  \ \ \ 
+3H\rho_{DE}
(1+w_{DE})=0.
\end{eqnarray}
Finally, substituting $H$ from (\ref{Hrel2}), and using   the dark energy density 
parameter definition (\ref{ODE}) we result to 
\begin{equation}\label{wDE}
w_{DE}=\frac{1-2\delta}{3}
-\frac{Q}{3}
(\Omega_{DE})^{\frac{1}{2(2-\delta) } } (1\!-\!\Omega_{DE})^{\frac{\delta-1}{
2(\delta-2) } }
e^{\frac{3(1-\delta)}{2(\delta-2)}x}.
\end{equation}
Thus, this expression provides $w_{DE}$ as a function of $\ln a$, as long as 
$\Omega_{DE}$ is known from the solution of (\ref{Odediffeq}). As expected, for 
$\delta=1$ 
(\ref{wDE}) does not have an explicit $x$-dependence and it gives the usual holographic 
dark energy equation-of-state parameter, namely 
$w_{DE}|_{\delta=1}=-\frac{1}{3}-\frac{2}{3}\sqrt{\frac{3M_p^2 \Omega_{DE}}{B}}$ 
\cite{Wang:2016och}, where complete coincidence is acquired under the identification  
$B=3 
c^2 M_p^2$.

We close this section  by introducing the convenient deceleration 
parameter $q$, which reads as
  \begin{equation}
  \label{qdeccel}
  q\equiv-1-\frac{\dot{H}}{H^2}=\frac{1}{2}+\frac{3}{2}\left(w_m\Omega_m+w_{DE}\Omega_{DE}
  \right).
\end{equation}
Hence, in the case of dust matter ($w_m=0$), $q$ is straightforwardly known as long as 
$\Omega_{DE}$ is known from  (\ref{Odediffeq}).

\section{Cosmological evolution}
\label{Cosmologicalevolution}

In this section we proceed to the investigation of the cosmological behavior in a 
universe where the dark energy sector is the Tsallis holographic dark energy. The basic 
differential equation that determines the evolution of  $\Omega_{DE}$ as a function of 
$x=\ln a$ is Eq. (\ref{Odediffeq}). Unfortunately, this equations can be analytically 
solved in an implicit form  only in the case $\delta=1$ \cite{Li:2004rb}, since in the 
$\delta\neq1$ case it acquires an explicit $x$-dependence that does not allow for an 
analytical solution. Hence, one should resort to numerical elaboration in order to 
extract its solution. As long as the solution for $\Omega_{DE}(x)$ is obtained, its 
behavior in terms of the redshift $z$ can be straightforwardly obtained   through 
$x\equiv\ln a=-\ln(1+z)$ (having set $a_0=1$). 

\begin{figure}[ht]
\includegraphics[scale=0.45]{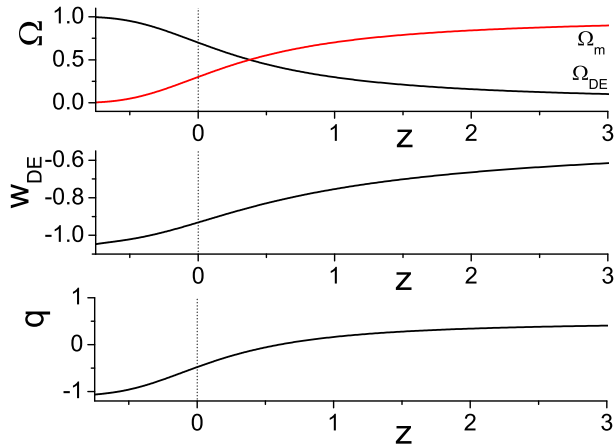}
\caption{
{\it{ Upper graph: The evolution of  Tsallis holographic dark energy    
density parameter $\Omega_{DE}$ (black-solid)  and of the matter density
parameter $\Omega_{m}$ (red-dashed), as a function of the redshift $z$,
for $\delta=1.1$ and $B=3$, in units where $M_p^2=1$. 
  Middle graph: The evolution of the corresponding dark
energy equation-of-state parameter $w_{DE}$. Lower graph:  The evolution of the 
corresponding   deceleration parameter $q$. In all graphs we have imposed 
 $\Omega_{DE}(x=-\ln(1+z)=0)\equiv\Omega_{DE0}\approx0.7$  at present in agreement with 
observations, and we have added a vertical dotted line denoting 
the present time $z=0$.
}} }
\label{Omegas}
\end{figure}

We elaborate  Eq. (\ref{Odediffeq}) numerically, imposing that 
$\Omega_{DE}(x=-\ln(1+z)=0)\equiv\Omega_{DE0}\approx0.7$ and thus 
$\Omega_m(x=-\ln(1+z)=0)\equiv\Omega_{m0}\approx0.3$  as required by observations 
\cite{Ade:2015xua}.  In the upper graph of Fig. \ref{Omegas} we present 
$\Omega_{DE}(z)$ and $\Omega_{m}(z)=1-\Omega_{DE}(z)$. In the middle graph we depict the 
corresponding behavior of $w_{DE}(z)$ as it arises from  (\ref{wDE}). And in the 
lower graph we draw the deceleration parameter from (\ref{qdeccel}).
Additionally, note that in the graphs we have extended the evolution into 
the future, namely for $z<0$, since $z\rightarrow-1$  corresponds to 
$t\rightarrow\infty$. 

From the upper graph of Fig. \ref{Omegas} we observe that we can acquire the usual 
thermal 
history of the universe, namely the sequence of matter and dark energy eras, while the 
universe asymptotically results in a complete dark-energy domination. Furthermore, from 
the third graph of Fig. \ref{Omegas} we can see that the transition 
from deceleration to acceleration happens at $z\approx 0.5$ as required from 
observations. Finally, from the middle graph of Fig. \ref{Omegas} we can see that the 
current value of $w_{DE}$ is around $-1$ in agreement with observations. We mention that 
in this explicit example in the future $w_{DE}$ enters slightly in the phantom regime, 
which according to relation (\ref{wDE}) is allowed in the model at hand, which is an 
advantage showing the enhanced capabilities.

\begin{figure}[ht]
\includegraphics[scale=0.45]{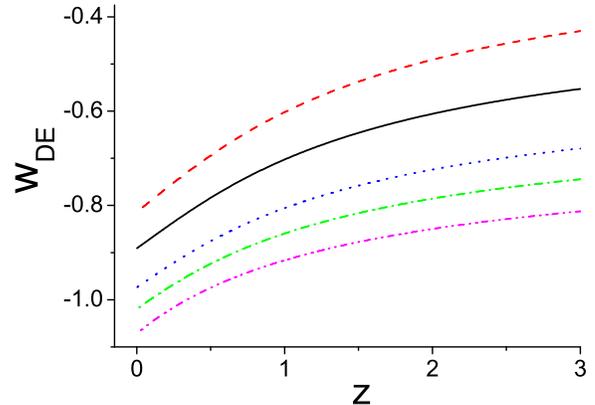}
\caption{
{\it{The evolution of the equation-of-state parameter $w_{DE}$ of Tsallis
holographic dark energy, as a function of the redshift $z$, for $B=3$, and  for 
$\delta=0.8 $ 
(red-dashed), $\delta=1$ (black-solid), $\delta=1.2$ 
(blue-dotted), and  $\delta=1.3$ (green-dashed-dotted), $\delta=1.4$ 
(magenta-dashed-dot-dotted), in units 
where $M_p^2=1$.
In all graphs we have imposed 
$\Omega_{DE}(x=-\ln(1+z)=0)\equiv\Omega_{DE0}\approx0.7$ at present in agreement with 
observations.}} }
\label{wzplot}
\end{figure}

Let us now investigate the effect of $\delta$ on $w_{DE}$. In Fig. \ref{wzplot} we 
present $w_{DE}(z)$ for various values of $\delta$, including the value $\delta=1$ which 
corresponds to standard holographic dark energy. As we observe, for increasing $\delta$ 
the $w_{DE}(z)$ evolution, as well as its present value $w_{DE}(z=0)$, tend to lower 
values. Note that for $\delta \gtrsim  1.2 $ the value of $w_{DE}(z=0)$ is in the 
phantom regime.  Hence, according to the value of $\delta$,  the dark energy sector can 
be quintessence-like, phantom-like, or experience  the phantom-divide crossing before or 
after the present time.

We mention here that although the scenario at hand has two parameters, namely the new 
exponent $\delta$ 
and the constant $B$, in the above examples we preferred to fix $B=3$, which is 
required 
in order to obtain exact coincidence with standard holographic dark energy when 
$\delta=1$, and explore the role of $\delta$ in a pure way. Nevertheless, as we showed, 
changing $\delta$ is adequate in order to obtain a cosmology in agreement with 
observations, without the need to change the constant $B$. This  is a significant 
advantage of Tsallis holographic dark energy comparing to standard 
holographic dark energy, since in the latter one needs to use a value of the constant 
$c^2$ different than 1 in order to obtain satisfying observational fittings, which has 
then difficulties to be theoretically justified (the essence of holographic dark energy 
is that the total energy in a region of size $L$ should not exceed the mass of a black
hole of the same size, and not the latter multiplied by a tuned constant).
Definitely, changing additionally the value of $B$ would significantly enhanced the 
capabilities of Tsallis holographic dark energy.

  \begin{figure*}[!]
  \includegraphics[width=0.8\textwidth]{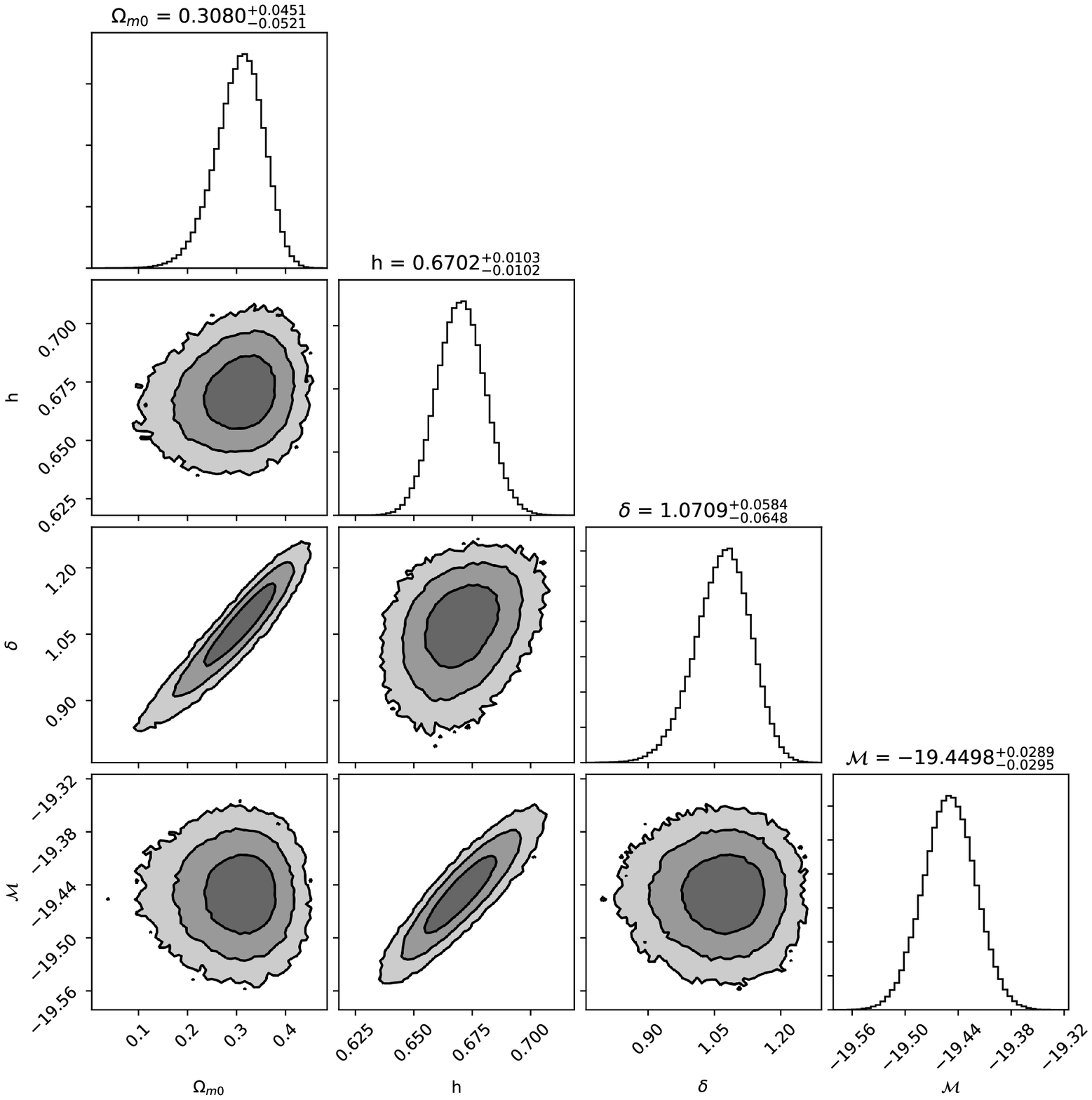}
\caption{
{\it{The $1\sigma$, $2\sigma$ and $3\sigma$ 2-dimensional contour 
plots for several combinations of various quantities of Tsallis holographic dark energy 
scenario, using SNIa and $H(z)$ data. Additionally, we depict the corresponding 
1-dimensional (1D) marginalized posterior distributions and the   mean
values of the parameters corresponding to the $1\sigma$ area of the MCMC chain. The 
parameter ${\cal{M}}$ is the usual free parameter of SNIa data that quantifies possible 
systematic errors of astrophysical origin \cite{Scolnic:2017caz}. For these fittings we 
obtain $\chi^2_{min}/dof = 43.248/76 $. }} }
\label{SNdata}
\end{figure*}

\section{Observational constraints}
\label{ObsCon}
 
 In this section we confront the scenario with cosmological data from Supernovae type 
Ia  observations as well as direct $H(z)$ Hubble data.
In particular, we desire to extract the  constraints on the free parameters of the model,
through  the maximum likelihood analysis. Assuming Gaussian errors this can be obtained 
by minimizing the $\chi^2$ function in terms of the model free parameters $a_m$. Since we 
utilize  SNIa and   Hublle rate    data, the total $\chi^2$ 
reads
\begin{equation}
    \chi^2_{tot} = \chi^2_{H} + \chi^2_{SN},
\end{equation}
where $\chi^2_{H}, \ \chi^2_{SN}$ will be defined in the following.
In the case of holographic dark energy through Tsallis entropy   the statistical vector of 
the 
free parameters is $a_m = (\Omega_{m0},B,\delta, h)$, with  
$h=H_{0}/100$. Note that, as we mentioned above,
we fix $B=3$, i.e. to its standard holographic dark energy value, in order to explore the 
role of $\delta$ in a pure way. We use the Markov Chain
Monte Carlo (MCMC) algorithm within the Python package emcee
\cite{ForemanMackey:2012ig}, in order to  minimize $\chi^2$ with respect to $a_m$. 
Finally, the algorithm convergence is checked with auto-correlation time considerations, 
while we 
also employ the Gelman-Rubin 
criterion \cite{Gelman:1992zz} for completeness.

\subsection{Type Ia Supernovae}

Type Ia Supernovae (SNIa) are widely used in cosmological confrontation, since they can 
be handled as ``standard candles'', offering  a way to measure cosmic
distances.  In these data sets one measures the apparent luminosity as function of 
redshift or equivalently the apparent 
magnitude. The $\chi^2$ that corresponds to the fit is given by
  \begin{equation}
  \chi ^2 _{SN} = \mu C_{SN,cov}^{-1} \mu^{T},
  \end{equation}
    where $\mu = \{ \mu_{\rm obs} - \mu_{\rm th}(z_{1}; a_m),..,\mu_{\rm obs} - \mu_{\rm 
th}(z_{N}; 
a_m) \} $ and $N=40$.  In the above expression
$\mu_{\rm obs}$ is the  observed distance modulus, which for every supernova is
defined as the difference between its absolute and 
apparent magnitude. We use the binned SNIa
data points and the corresponding  inverse covariance matrix  $C_{SN,cov}^{-1}$ from 
\cite{Scolnic:2017caz}.
Moreover, in the statistical vector $a_m$ we include  the quantity ${\cal{M}}$ that 
quantifies errors  of astrophysical origins in 
the observed distance moduli.
% considered to be uncorrelated and gaussian.
The theoretically calculated 
distance modulus $\mu_{\rm th}$  has a dependence on the model parameters
$a_m$  which reads as
\begin{equation}\mu_{\rm th}\left(z\right)=42.38-5\log_{10}h+5\log_{10}\[D_{L}
\left(z;a_m\right)\],
  \end{equation}
where the dimensionless luminosity distance $D_{L}(z;a_m)$ is given by
  \begin{equation} D_{L}\(z;a_m\)\equiv\left(1+z\right)
\int^{z}_{0}dz'\frac{H_0}{H\left(z';a_m\right)}.
  \end{equation}
  Note that the quantity $H\left(z';a_m\right)$ in the 
scenario at hand cannot be obtained analytically and thus it is extracted from  
(\ref{Hrel2}) after the    numerical elaboration of Eq. (\ref{Odediffeq}).

\subsection{Cosmic chronometer Hubble data}
\label{cc-data}

The direct measurements of the Hubble constant is a very powerful implementation in 
cosmological confrontation, introduced first in \cite{Jimenez:2001gg} with the method 
determining the Hubble parameter data through the differential age evolution of the 
passively evolving early-type galaxies. In particular, since   the Hubble function    can 
be expressed as $H= -(1+z)^{-1} dz/dt$, by measuring $dz/dt$ one can 
directly measure $H(z)$ data (see  \cite{Moresco11} for a detailed discussion).

In this work we use the most recent Hubble data  from \cite{Yu:2017iju}. The corresponding
dataset contains $N=36$ measurements of $H(z)$ in  the $0.07\leq z\leq 2.33$
range. Amongst these, there are 5 data points that are based on Baryon Acoustic 
Oscillations (BAO), nevertheless for all the remaining points   the Hubble 
constant is measured through the passive evolving galaxies differential age. The $\chi^2$ 
that corresponds to the fit is given by
\begin{equation}
\chi^{2}_{H}\left(a_m\right)={\cal H}\,
C_{H,\text{cov}}^{-1}\,{\cal H}^{T}\,,
\end{equation}
where ${\cal H}=\{H_{1}-H_{0}E(z_{1},a_m)\,,\,...\,,\,
H_{N}-H_{0}E(z_{N},a_m)\}$, $H_{i}$ is  the observed Hubble values at 
redshifts $z_{i}$ ($i=1,...,N$), and with  $C$ being the  
involved covariance matrix (see  \cite{Nunes:2016qyp,Basilakos:2018arq} for more 
details). 
Note that the theoretical quantity $E(z_{i},a_m)\equiv 
H(z_{i},a_m)/H_0$ in the scenario at hand has to be extracted numerically through 
  (\ref{Hrel2})  and
(\ref{Odediffeq}).

In Fig.  \ref{SNdata} we present the  contour 
plots for several combinations of various quantities of Tsallis holographic dark energy 
scenario, using SNIa and $H(z)$ data. Additionally, we depict the corresponding 
1-dimensional (1D) marginalized posterior distributions and the   mean
values of the parameters corresponding to the $1\sigma$ area of the MCMC chain. 
 As we can see the agreement with the data is very good, and the matter energy density as 
well as the current value of the Hubble parameter coincide with those of Planck within 
1$\sigma$ \cite{Ade:2015xua}. Concerning the new physical 
parameter of the present work, namely the exponent $\delta$, there is a tendency for a 
slight  deviation from its standard value 1, however the value 1 is included within 
2$\sigma$.
 Moreover, note that these results could be improved allowing 
$B$ to change too (in this case in the end one has to use the AIC \cite{Akaike1974} and 
BIC \cite{bic} criteria in order to 
consistently weight the effect of the additional model parameters). We mention here 
that the incorporation of other datasets such as Cosmic Microwave Background (CMB), 
although necessary, would require a special treatment of the   $H(z)$ form, which in the 
current scenario cannot be obtained analytically in general. This complicated elaboration 
 lies 
beyond the scope of the present work and is left for future investigation, along with the 
perturbation analysis and the use of Large Scale Structure data.

\section{Conclusions}
\label{Conclusions}

In the present work we formulated Tsallis holographic dark energy, which is a 
generalization of standard holographic dark energy.  In particular, in order to apply 
holography and entropy relations to the whole universe, which is a gravitational and thus
nonextensive system, for consistency one should use the generalized definition of the 
universe horizon entropy, namely Tsallis nonextensive entropy, quantified by a new 
dimensionless parameter $\delta$. Although a similar idea appeared in a recent work too 
\cite{Tavayef:2018xwx}, its cosmological application had the serious disadvantage that it 
did not possess standard entropy and standard holographic dark energy as a sub-case, due 
to the fact that it was the Hubble horizon that was used as the IR  cutoff, which is well 
known that cannot lead to realistic cosmology in case of usual holographic dark energy. 
On 
the other hand, in the present investigation we presented a consistent formulation of 
Tsallis holographic dark energy, taking  the IR cutoff to be the future event horizon, 
namely the same length that is used in standard  holographic dark energy scenario. In 
this 
way Tsallis holographic dark energy is indeed a consistent generalization of standard 
holographic dark energy, possessing it as a particular limit, namely for $\delta=1$.

In order to study the cosmological applications of Tsallis holographic dark energy we 
first provided a simple  differential equation for the holographic dark energy density 
parameter $\Omega_{DE}$. Additionally, we extracted an analytical expression for the 
holographic dark energy equation-of-state parameter $w_{DE}$ as a function of 
$\Omega_{DE}$. Although in the case  $\delta=1$ the above differential equation can be 
solved analytically in an implicit form, in the general case it does not accept an 
analytical solution and thus one has to elaborate it numerically.

The scenario of Tsallis holographic dark energy leads to interesting cosmological 
phenomenology. Firstly, the universe exhibits the usual thermal history, namely the 
successive sequence of matter and dark-energy epochs, with the transition from 
deceleration to acceleration happening at $z\approx0.5$ in agreement with observations, 
before it results in a complete dark energy domination in the far future. Furthermore, 
the corresponding dark energy equation-of-state parameter presents a rich behavior, and
according to the value of $\delta$, it can be quintessence-like, phantom-like, or 
experience the phantom-divide crossing before or after the present time. 

 Additionally, we 
confronted the scenario with Supernovae type Ia  and $H(z)$ observational data, 
we constructed 
the corresponding contour plots, and we saw that the 
agreement is very good. Concerning the new physical 
parameter of the present work, namely the exponent $\delta$, there is a tendency for a 
slight  deviation from its standard value 1, however the value 1 is included within 
2$\sigma$. 

We mention that the above behaviors were obtained changing only the value of $\delta$ and 
keeping the second parameter (the one that is present in holographic dark energy models) 
$B$ fixed. This  is a significant advantage   comparing to standard 
holographic dark energy, since in the latter one needs to use a value of this parameter 
different than the straightforward one in order to obtain satisfying observational 
fittings, which has then difficulties to be theoretically justified.
Definitely, changing additionally the value of $B$ enhances significantly the 
capabilities of Tsallis holographic dark energy.

In summary, as we can see, the scenario of Tsallis holographic dark energy exhibits 
richer behavior comparing to standard holographic dark energy, quantified by the present 
of the new parameter $\delta$, while due to its consistent formulation one can still 
obtain as a sub-case the scenario of standard holographic dark energy, namely for 
$\delta=1$. There are  additional studies that need to be performed before  the scenario  
can be considered as a successful candidate for the description of nature.   Firstly, 
one should perform a joint observational analysis at both the background and 
perturbation levels, using  data from Cosmic Microwave Background (CMB) and Large Scale 
Structure (such as f$\sigma$8), in order to constrain the model parameters.  Moreover, 
one should perform a detailed phase-space  analysis in order to extract the global 
features of the 
scenario at late times, independently of the initial conditions and the specific 
evolution. These necessary investigations lie beyond the scope of this work and are 
left for future projects.

\section*{Acknowledgments}
 The authors wish to thank C. Tsallis and K. Ntrekis for useful 
discussions. The work of KB is supported in part by the JSPS KAKENHI Grant Number JP 
25800136 and Competitive Research Funds for Fukushima University Faculty 
(17RI017). This article is based upon 
work from COST Action CA15117 ``Cosmology and Astrophysics Network for Theoretical 
Advances and Training Actions'' (CANTATA), supported by COST (European
Cooperation in Science and Technology).


\begin{thebibliography}{99}

 
 
   
  
 
  
 
%\cite{Copeland:2006wr}
\bibitem{Copeland:2006wr}
  E.~J.~Copeland, M.~Sami and S.~Tsujikawa,
%     {\it{Dynamics of dark energy}},
      \href{\doi/10.1142/S021827180600942X}{Int.\ J.\ Mod.\ Phys.\  D {\bf 15}, 1753 
(2006)}
  [\href{\arxiv/hep-th/0603057}{hep-th/0603057}].
 
 
 

%\cite{Cai:2009zp}
\bibitem{Cai:2009zp}
  Y.~F.~Cai, E.~N.~Saridakis, M.~R.~Setare and J.~Q.~Xia,
%     {\it{Quintom Cosmology: Theoretical implications and observations}},
      \href{\doi/10.1016/j.physrep.2010.04.001}{Phys.\ Rept.\  {\bf 493}, 1 (2010)}
  [\href{\arxiv/arXiv:0909.2776}{0909.2776} [hep-th]].
 
 
 %\cite{Bamba:2012cp}
\bibitem{Bamba:2012cp} 
  K.~Bamba, S.~Capozziello, S.~Nojiri and S.~D.~Odintsov,
  %``Dark energy cosmology: the equivalent description via different theoretical models 
%and cosmography tests,''
      \href{\doi/10.1016/10.1007/s10509-012-1181-8}{Astrophys.\ Space Sci.\  {\bf 342}, 
155 (2012)}
  [\href{\arxiv/arXiv:1205.3421}{1205.3421} [gr-qc]].

 
  
  

%\cite{Nojiri:2006ri}
\bibitem{Nojiri:2006ri}
  S.~'i.~Nojiri and S.~D.~Odintsov,
 % {\it{Introduction to modified gravity and gravitational alternative for
%dark energy}},
  eConf C {\bf 0602061}, 06 (2006)
        \href{\doi/10.1142/S0219887807001928}{Int.\ J.\ Geom.\ Meth.\ Mod.\ Phys.\  {\bf 
4}, 115 (2007)}
  [\href{\arxiv/hep-th/0601213}{hep-th/0601213}].
 
 
 \bibitem{Nojiri:2010wj}
 S.~Nojiri and S.~D.~Odintsov,
 %``Unified cosmic history in modified gravity: from F(R) theory to 
%Lorentz non-invariant models,''
       \href{\doi/10.1016/10.1016/j.physrep.2011.04.001}{Phys.\ Rept.\ {\bf 505}, 59 
(2011)}
  [\href{\arxiv/arXiv:1011.0544}{1011.0544} [gr-qc]].
 
 
 
  
%\cite{Capozziello:2011et}
\bibitem{Capozziello:2011et}
  S.~Capozziello and M.~De Laurentis,
 % {\it{Extended Theories of Gravity}},
       \href{\doi/10.1016/j.physrep.2011.09.003}{Phys.\ Rept.\  {\bf 509}, 167 (2011)}
  [\href{\arxiv/arXiv:1108.6266}{1108.6266} [gr-qc]].
 
 
 
 
  
%\cite{Cai:2015emx}
\bibitem{Cai:2015emx} 
  Y.~F.~Cai, S.~Capozziello, M.~De Laurentis and E.~N.~Saridakis,
  %``f(T) teleparallel gravity and cosmology,''
         \href{\doi/10.1088/0034-4885/79/10/106901}{Rept.\ Prog.\ Phys.\  {\bf 79}, no. 
10, 106901 (2016)}
  [\href{\arxiv/arXiv:1511.07586}{1511.07586} [gr-qc]].
 
 %\cite{Nojiri:2017ncd}
\bibitem{Nojiri:2017ncd} 
  S.~Nojiri, S.~D.~Odintsov and V.~K.~Oikonomou,
  %``Modified Gravity Theories on a Nutshell: Inflation, Bounce and Late-time Evolution,''
         \href{\doi/10.1016/j.physrep.2017.06.001}{Phys.\ Rept.\  {\bf 692}, 1 (2017)}
  [\href{\arxiv/arXiv:1705.11098}{1705.11098} [gr-qc]].
 
 
  
 
   
%\cite{tHooft:1993dmi}
\bibitem{tHooft:1993dmi} 
  G.~'t Hooft,
  %``Dimensional reduction in quantum gravity,''
          Salamfest 1993: 0284-296
  [\href{\arxiv/gr-qc/9310026}{gr-qc/9310026}].
 
 
 
  %\cite{Susskind:1994vu}
\bibitem{Susskind:1994vu} 
  L.~Susskind,
  %``The World as a hologram,''
          \href{\doi/10.1063/1.531249}{ J.\ Math.\ Phys.\  {\bf 36}, 6377 (1995)}
  [\href{\arxiv/hep-th/9409089}{hep-th/9409089}].
 
   
  
  %\cite{Bousso:2002ju}
\bibitem{Bousso:2002ju} 
  R.~Bousso,
  %``The Holographic principle,''
          \href{\doi/10.1103/RevModPhys.74.825}{Rev.\ Mod.\ Phys.\  {\bf 74}, 825 (2002)}
  [\href{\arxiv/hep-th/0203101}{hep-th/0203101}].
 
   
  %\cite{Fischler:1998st}
\bibitem{Fischler:1998st} 
  W.~Fischler and L.~Susskind,
  %``Holography and cosmology,''
    [\href{\arxiv/hep-th/9806039}{hep-th/9806039}].
 
   
  
  %\cite{Bak:1999hd}
\bibitem{Bak:1999hd} 
  D.~Bak and S.~J.~Rey,
  %``Cosmic holography,''
            \href{\doi/10.1088/0264-9381/17/15/101}{ Class.\ Quant.\ Grav.\  {\bf 17}, 
L83 (2000)}
  [\href{\arxiv/hep-th/9902173}{hep-th/9902173}].
 
  
    
  
  %\cite{Horava:2000tb}
\bibitem{Horava:2000tb} 
  P.~Horava and D.~Minic,
  %``Probable values of the cosmological constant in a holographic theory,''
  \href{\doi/10.1103/PhysRevLett.85.1610}{Phys.\ Rev.\ Lett.\  {\bf 85}, 1610 
(2000)}
  [\href{\arxiv/hep-th/0001145}{hep-th/0001145}].
 
  
     
  
  
%\cite{Cohen:1998zx}
\bibitem{Cohen:1998zx} 
  A.~G.~Cohen, D.~B.~Kaplan and A.~E.~Nelson,
  %``Effective field theory, black holes, and the cosmological constant,''
    \href{\doi/10.1103/PhysRevLett.82.4971}{Phys.\ Rev.\ Lett.\  {\bf 82}, 4971 (1999)}
  [\href{\arxiv/hep-th/9803132}{hep-th/9803132}].
 
  
  
  
  
  
    %\cite{Li:2004rb}
\bibitem{Li:2004rb} 
  M.~Li,
  %``A Model of holographic dark energy,''
      \href{\doi/10.1016/j.physletb.2004.10.014}{Phys.\ Lett.\ B {\bf 603}, 1 (2004)}
  [\href{\arxiv/hep-th/0403127}{hep-th/0403127}].
  
  
  
  %\cite{Wang:2016och}
\bibitem{Wang:2016och} 
  S.~Wang, Y.~Wang and M.~Li,
  %``Holographic Dark Energy,''
        \href{\doi/10.1016/j.physrep.2017.06.003}{Phys.\ Rept.\  {\bf 696}, 1 (2017)}
  [\href{\arxiv/arXiv:1612.00345}{1612.00345} [astro-ph.CO]].
 
 
 
 
  
  
  
   
  
  %\cite{Horvat:2004vn}
\bibitem{Horvat:2004vn} 
  R.~Horvat,
  %``Holography and variable cosmological constant,''
       \href{\doi/10.1103/PhysRevD.70.087301}{ Phys.\ Rev.\ D {\bf 70}, 087301 (2004)}
  [\href{\arxiv/astro-ph/0404204}{astro-ph/0404204}].
  
 
  
    
  %\cite{Huang:2004ai}
\bibitem{Huang:2004ai} 
  Q.~G.~Huang and M.~Li,
  %``The Holographic dark energy in a non-flat universe,''
            \href{\doi/10.1088/1475-7516/2004/08/013}{ JCAP {\bf 0408}, 013 (2004)}
  [\href{\arxiv/astro-ph/0404229}{astro-ph/0404229}].
  
 
%\cite{Pavon:2005yx}
\bibitem{Pavon:2005yx} 
  D.~Pavon and W.~Zimdahl,
  %``Holographic dark energy and cosmic coincidence,''
 \href{\doi/10.1016/j.physletb.2005.08.134}{Phys.\ Lett.\ B {\bf 628}, 206 
(2005)}
  [\href{\arxiv/gr-qc/0505020}{gr-qc/0505020}].
  
  
   
    %\cite{Wang:2005jx}
\bibitem{Wang:2005jx} 
  B.~Wang, Y.~g.~Gong and E.~Abdalla,
  %``Transition of the dark energy equation of state in an interacting holographic dark 
%energy model,''
          \href{\doi/10.1016/j.physletb.2005.08.008}{  Phys.\ Lett.\ B {\bf 624}, 141 
(2005)}
  [\href{\arxiv/hep-th/0506069}{hep-th/0506069}].
  
  
   

%\cite{Nojiri:2005pu}
\bibitem{Nojiri:2005pu} 
  S.~Nojiri and S.~D.~Odintsov,
  %``Unifying phantom inflation with late-time acceleration: Scalar phantom-non-phantom 
%transition model and generalized holographic dark energy,''
          \href{\doi/10.1007/s10714-006-0301-6}{ Gen.\ Rel.\ Grav.\  {\bf 38}, 1285 
(2006)}
  [\href{\arxiv/hep-th/0506212}{hep-th/0506212}].
   


    %\cite{Kim:2005at}
\bibitem{Kim:2005at} 
  H.~Kim, H.~W.~Lee and Y.~S.~Myung,
  %``Equation of state for an interacting holographic dark energy model,''
   \href{\doi/10.1016/j.physletb.2005.11.043}{Phys.\ Lett.\ B {\bf 632}, 605 (2006)}
  [\href{\arxiv/gr-qc/0509040}{gr-qc/0509040}].
  
   
  

  %\cite{Wang:2005ph}
\bibitem{Wang:2005ph} 
  B.~Wang, C.~Y.~Lin and E.~Abdalla,
  %``Constraints on the interacting holographic dark energy model,''
            \href{\doi/10.1016/j.physletb.2006.04.009}{ Phys.\ Lett.\ B {\bf 637}, 357 
(2006)}
  [\href{\arxiv/hep-th/0509107}{hep-th/0509107}].
  
   
  %\cite{Setare:2006wh}
\bibitem{Setare:2006wh} 
  M.~R.~Setare,
  %``Interacting holographic dark energy model in non-flat universe,''
 \href{\doi/10.1016/j.physletb.2006.09.027}{Phys.\ Lett.\ B {\bf 642}, 1 
(2006)}
  [\href{\arxiv/hep-th/0609069}{hep-th/0609069}].
  
   
  %\cite{Setare:2008pc}
\bibitem{Setare:2008pc} 
  M.~R.~Setare and E.~N.~Saridakis,
  %``Non-minimally coupled canonical, phantom and quintom models of holographic dark 
%energy,''
         \href{\doi/10.1016/j.physletb.2008.12.026}{Phys.\ Lett.\ B {\bf 671}, 331 (2009)}
  [\href{\arxiv/arXiv:0810.0645}{0810.0645} [hep-th]].
  
   
  %\cite{Setare:2008hm}
\bibitem{Setare:2008hm} 
  M.~R.~Setare and E.~N.~Saridakis,
  %``Correspondence between Holographic and Gauss-Bonnet dark energy models,''
           \href{\doi/10.1016/j.physletb.2008.10.029}{Phys.\ Lett.\ B {\bf 670}, 1 (2008)}
  [\href{\arxiv/arXiv:0810.3296}{0810.3296} [hep-th]].
  
    
  
  %\cite{Gong:2004fq}
\bibitem{Gong:2004fq} 
  Y.~g.~Gong,
  %``Extended holographic dark energy,''
   \href{\doi/10.1103/PhysRevD.70.064029}{ Phys.\ Rev.\ D {\bf 70}, 064029 (2004)}
  [\href{\arxiv/hep-th/0404030}{hep-th/0404030}].
  
  
    
%\cite{Saridakis:2007cy}
\bibitem{Saridakis:2007cy} 
  E.~N.~Saridakis,
  %``Restoring holographic dark energy in brane cosmology,''
 \href{\doi/10.1016/j.physletb.2008.01.004}{Phys.\ Lett.\ B {\bf 660}, 138 (2008)}
  [\href{\arxiv/arXiv:0712.2228}{0712.2228} [hep-th]].
  
    
     %\cite{Setare:2007we}
\bibitem{Setare:2007we} 
  M.~R.~Setare and E.~C.~Vagenas,
  %``The Cosmological dynamics of interacting holographic dark energy model,''
    \href{\doi/10.1142/S0218271809014303}{ Int.\ J.\ Mod.\ Phys.\ D {\bf 18}, 147 (2009)}
  [\href{\arxiv/arXiv:0704.2070}{0704.2070} [hep-th]].
  
  
  
  %\cite{Cai:2007us}
\bibitem{Cai:2007us} 
  R.~G.~Cai,
  %``A Dark Energy Model Characterized by the Age of the Universe,''
    \href{\doi/10.1016/j.physletb.2007.09.061}{Phys.\ Lett.\ B {\bf 657}, 228 (2007)}
  [\href{\arxiv/arXiv:0707.4049}{0707.4049} [hep-th]].
   
   


  
  %\cite{Saridakis:2007ns}
\bibitem{Saridakis:2007ns} 
  E.~N.~Saridakis,
  %``Holographic Dark Energy in Braneworld Models with Moving Branes and the w=-1 
%Crossing,''
 \href{\doi/10.1088/1475-7516/2008/04/020}{ JCAP {\bf 0804}, 020 (2008)}
  [\href{\arxiv/arXiv:0712.2672}{0712.2672} [astro-ph]].
  
   
  
  %\cite{Saridakis:2007wx}
\bibitem{Saridakis:2007wx} 
  E.~N.~Saridakis,
  %``Holographic Dark Energy in Braneworld Models with a Gauss-Bonnet Term in the Bulk. 
%Interacting Behavior and the w =-1 Crossing,''
             \href{\doi/10.1016/j.physletb.2008.02.032}{Phys.\ Lett.\ B {\bf 661}, 335 
(2008)}
  [\href{\arxiv/arXiv:0712.3806}{0712.3806} [gr-qc]].
  
     
      
    %\cite{Setare:2008bb}
\bibitem{Setare:2008bb} 
  M.~R.~Setare and E.~C.~Vagenas,
  %``Thermodynamical Interpretation of the Interacting Holographic Dark Energy Model in a 
%non-flat Universe,''
   \href{\doi/10.1016/j.physletb.2008.07.013}{Phys.\ Lett.\ B {\bf 666}, 111 (2008)}
  [\href{\arxiv/arXiv:0801.4478}{0801.4478} [hep-th]].
 
  
%\cite{Jamil:2009sq}
\bibitem{Jamil:2009sq} 
  M.~Jamil, E.~N.~Saridakis and M.~R.~Setare,
  %``Holographic dark energy with varying gravitational constant,''
   \href{\doi/10.1016/j.physletb.2009.07.048}{Phys.\ Lett.\ B {\bf 679}, 172 (2009)}
  [\href{\arxiv/arXiv:0906.2847}{0906.2847} [hep-th]].
  
      
  %\cite{Gong:2009dc}
\bibitem{Gong:2009dc} 
  Y.~Gong and T.~Li,
  %``A Modified Holographic Dark Energy Model with Infrared Infinite Extra Dimension(s),''
     \href{\doi/10.1016/j.physletb.2009.12.040}{Phys.\ Lett.\ B {\bf 683}, 241 (2010)}
  [\href{\arxiv/arXiv:0907.0860}{0907.0860} [hep-th]].
  
   
  
    
    %\cite{Suwa:2009gm}
\bibitem{Suwa:2009gm} 
  M.~Suwa and T.~Nihei,
  %``Observational constraints on the interacting Ricci dark energy model,''
             \href{\doi/10.1103/PhysRevD.81.023519}{Phys.\ Rev.\ D {\bf 81}, 023519 
(2010)}
  [\href{\arxiv/arXiv:0911.4810}{0911.4810} [astro-ph.CO]].
  
     
  
    %\cite{Jamil:2010vr}
\bibitem{Jamil:2010vr} 
  M.~Jamil and E.~N.~Saridakis,
  %``New agegraphic dark energy in Horava-Lifshitz cosmology,''
        \href{\doi/10.1088/1475-7516/2010/07/028}{JCAP {\bf 1007}, 028 (2010)}
  [\href{\arxiv/arXiv:1003.5637}{1003.5637} [hep-th]].
   
    
    
  
 
  
    
  %\cite{BouhmadiLopez:2011xi}
\bibitem{BouhmadiLopez:2011xi} 
  M.~Bouhmadi-Lopez, A.~Errahmani and T.~Ouali,
  %``The cosmology of an holographic induced gravity model with curvature effects,''
       \href{\doi/10.1103/PhysRevD.84.083508}{Phys.\ Rev.\ D {\bf 84}, 083508 (2011)}
  [\href{\arxiv/arXiv:1104.1181}{1104.1181} [astro-ph.CO]].
  
    

    %\cite{Malekjani:2012bw}
\bibitem{Malekjani:2012bw} 
  M.~Malekjani,
  %``Generalized holographic dark energy model described at the Hubble length,''
   \href{\doi/10.1007/s10509-013-1522-2}{Astrophys.\ Space Sci.\  {\bf 347}, 405 
(2013)}
  [\href{\arxiv/arXiv:1209.5512}{1209.5512} [gr-qc]].
  
    
  
  %\cite{Khurshudyan:2014axa}
\bibitem{Khurshudyan:2014axa} 
  M.~Khurshudyan, J.~Sadeghi, R.~Myrzakulov, A.~Pasqua and H.~Farahani,
  %``Interacting quintessence dark energy models in Lyra manifold,''
     \href{\doi/10.1155/2014/878092}{Adv.\ High Energy Phys.\  {\bf 2014}, 878092 
(2014)}
  [\href{\arxiv/arXiv:1404.2141}{1404.2141} [gr-qc]].
  
    
    
    %\cite{Landim:2015hqa}
\bibitem{Landim:2015hqa} 
  R.~C.~G.~Landim,
  %``Holographic dark energy from minimal supergravity,''
       \href{\doi/10.1142/S0218271816500504}{ Int.\ J.\ Mod.\ Phys.\ D {\bf 25}, no. 04, 
1650050 (2016)}
  [\href{\arxiv/arXiv:1508.07248}{1508.07248} [hep-th]].
   
    %\cite{Pasqua:2015bfz}
\bibitem{Pasqua:2015bfz} 
  A.~Pasqua, S.~Chattopadhyay and R.~Myrzakulov,
  %``Power-law entropy-corrected holographic dark energy in Hořava-Lifshitz cosmology 
with 
%Granda-Oliveros cut-off,''
  \href{\doi/10.1140/epjp/i2016-16408-8}{  Eur.\ Phys.\ J.\ Plus {\bf 131}, no. 11, 408 
(2016)}
  [\href{\arxiv/arXiv:1511.00611}{1511.00611} [gr-qc]].
   
   
   
  
  

  

    
%\cite{Jawad:2016tne}
\bibitem{Jawad:2016tne} 
  A.~Jawad, N.~Azhar and S.~Rani,
  %``Entropy corrected holographic dark energy models in modified gravity,''
   \href{\doi/10.1142/S0218271817500407}{Int.\ J.\ Mod.\ Phys.\ D {\bf 26}, no. 04, 
1750040 (2016)}.
 
  %\cite{Pourhassan:2017cba}
\bibitem{Pourhassan:2017cba} 
  B.~Pourhassan, A.~Bonilla, M.~Faizal and E.~M.~C.~Abreu,
  %``Holographic Dark Energy from Fluid/Gravity Duality Constraint by Cosmological 
%Observations,''
  \href{\doi/10.1016/j.dark.2018.02.006}{Phys.\ Dark Univ.\  {\bf 20}, 41 (2018)}
  [\href{\arxiv/arXiv:1704.03281}{1704.03281} [hep-th]].
  
  
  
    %\cite{Nojiri:2017opc}
\bibitem{Nojiri:2017opc} 
  S.~Nojiri and S.~D.~Odintsov,
   \href{\doi/10.1140/epjc/s10052-017-5097-x}{  Eur.\ Phys.\ J.\ C  {\bf 77}, no. 8, 528 
(2017)}  
  %``Covariant Generalized Holographic Dark Energy and Accelerating Universe,''
    [\href{\arxiv/arXiv:1703.06372}{1703.06372} [hep-th]].
  
  
  
  %\cite{Saridakis:2017rdo}
\bibitem{Saridakis:2017rdo} 
  E.~N.~Saridakis,
  %``Ricci-Gauss-Bonnet holographic dark energy,''
         \href{\doi/10.1103/PhysRevD.97.064035}{Phys.\ Rev.\ D {\bf 97}, no. 6, 064035 
(2018)}
  [\href{\arxiv/arXiv:1707.09331}{1707.09331} [gr-qc]].
  
    
  
 
  %\cite{Zhang:2005hs}
\bibitem{Zhang:2005hs} 
  X.~Zhang and F.~Q.~Wu,
  %``Constraints on holographic dark energy from Type Ia supernova observations,''
       \href{\doi/10.1103/PhysRevD.72.043524}{Phys.\ Rev.\ D {\bf 72}, 043524 (2005)}
  [\href{\arxiv/astro-ph/0506310}{astro-ph/0506310}].
  
     
  
  %\cite{Li:2009bn}
\bibitem{Li:2009bn} 
  M.~Li, X.~D.~Li, S.~Wang and X.~Zhang,
  %``Holographic dark energy models: A comparison from the latest observational data,''
         \href{\doi/10.1088/1475-7516/2009/06/036}{JCAP {\bf 0906}, 036 (2009)}
  [\href{\arxiv/arXiv:0904.0928}{0904.0928} [astro-ph.CO]].
  
     

%\cite{Feng:2007wn}
\bibitem{Feng:2007wn} 
  C.~Feng, B.~Wang, Y.~Gong and R.~K.~Su,
  %``Testing the viability of the interacting holographic dark energy model by using 
%combined observational constraints,''
 \href{\doi/10.1088/1475-7516/2007/09/005}{JCAP {\bf 0709}, 005 (2007)}
  [\href{\arxiv/arXiv:0706.4033}{0706.4033} [astro-ph]].
  
   
%\cite{Zhang:2009un}
\bibitem{Zhang:2009un} 
  X.~Zhang,
  %``Holographic Ricci dark energy: Current observational constraints, quintom feature, 
%and the reconstruction of scalar-field dark energy,''
         \href{\doi/10.1103/PhysRevD.79.103509}{  Phys.\ Rev.\ D {\bf 79}, 103509 
(2009)}
  [\href{\arxiv/arXiv:0901.2262}{0901.2262} [astro-ph.CO]].
  
     
%\cite{Lu:2009iv}
\bibitem{Lu:2009iv} 
  J.~Lu, E.~N.~Saridakis, M.~R.~Setare and L.~Xu,
  %``Observational constraints on holographic dark energy with varying gravitational 
%constant,''
         \href{\doi/10.1088/1475-7516/2010/03/031}{JCAP {\bf 1003}, 031 (2010)}
  [\href{\arxiv/arXiv:0912.0923}{0912.0923} [astro-ph.CO]].
  
   

%\cite{Micheletti:2009jy}
\bibitem{Micheletti:2009jy} 
  S.~M.~R.~Micheletti,
  %``Observational constraints on holographic tachyonic dark energy in interaction with 
%dark matter,''
     \href{\doi/10.1088/1475-7516/2010/05/009}{JCAP {\bf 1005}, 009 (2010)}
  [\href{\arxiv/arXiv:0912.3992}{0912.3992} [gr-qc]].
  
    
      %\cite{Tsallis:1987eu}
\bibitem{Tsallis:1987eu} 
  C.~Tsallis,
  %``Possible Generalization of Boltzmann-Gibbs Statistics,''
       \href{\doi/10.1007/BF01016429}{ J.\ Statist.\ Phys.\  {\bf 52}, 479 (1988)}.
 

  
  %\cite{Lyra:1998wz}
\bibitem{Lyra:1998wz} 
  M.~L.~Lyra and C.~Tsallis,
  %``Nonextensivity and multifractality in low dissipative systems,''
       \href{\doi/10.1103/PhysRevLett.80.53}{ Phys.\ Rev.\ Lett.\  {\bf 80}, 53 (1998)}.
   
%\cite{Tsallis:1998ws}
\bibitem{Tsallis:1998ws} 
  C.~Tsallis, R.~S.~Mendes and A.~R.~Plastino,
  %``The Role of constraints within generalized nonextensive statistics,''
         \href{\doi/10.1016/S0378-4371(98)00437-3}{Physica A {\bf 261}, 534 (1998)}.
   
   
 
  
  
  
  
%\cite{Wilk:1999dr}
\bibitem{Wilk:1999dr} 
  G.~Wilk and Z.~Wlodarczyk,
  %``On the interpretation of nonextensive parameter q in Tsallis statistics and Levy 
%distributions,''
     \href{\doi/10.1103/PhysRevLett.84.2770}{Phys.\ Rev.\ Lett.\  {\bf 84}, 2770 (2000)}
  [\href{\arxiv/arXiv:hep-ph/9908459}{hep-ph/9908459}].
  
    
    
   
  
  
  
  %\cite{Tsallis:2012js}
\bibitem{Tsallis:2012js} 
  C.~Tsallis and L.~J.~L.~Cirto,
  %``Black hole thermodynamical entropy,''
       \href{\doi/10.1140/epjc/s10052-013-2487-6}{Eur.\ Phys.\ J.\ C {\bf 73}, 2487 
(2013)}
  [\href{\arxiv/arXiv:1202.2154}{1202.2154}[cond-mat.stat-mech]].
  
    
     
  


  
  %\cite{Tavayef:2018xwx}
\bibitem{Tavayef:2018xwx} 
  M.~Tavayef, A.~Sheykhi, K.~Bamba and H.~Moradpour,
  %``Tsallis Holographic Dark Energy,''
       \href{\doi/10.1016/j.physletb.2018.04.001}{Phys.\ Lett.\ B {\bf 781}, 195 (2018)}
  [\href{\arxiv/arXiv:1804.02983}{1804.02983} [gr-qc]].
  
  
   
    %\cite{Jahromi:2018xxh}
\bibitem{Jahromi:2018xxh} 
  A.~Sayahian Jahromi, S.~A.~Moosavi, H.~Moradpour, J.~P.~Morais Graça, I.~P.~Lobo, 
I.~G.~Salako and A.~Jawad,
  %``Generalized entropy formalism and a new holographic dark energy model,''
      \href{\doi/10.1016/j.physletb.2018.02.052}{Phys.\ Lett.\ B {\bf 780}, 21 (2018)}
  [\href{\arxiv/arXiv:1802.07722}{1802.07722} [gr-qc]].
  
  
  
   
  
  %\cite{Moradpour:2018ivi}
\bibitem{Moradpour:2018ivi} 
  H.~Moradpour, S.~A.~Moosavi, I.~P.~Lobo, J.~P.~Morais Graça, A.~Jawad and I.~G.~Salako,
  %``Thermodynamic approach to holographic dark energy and the R\'{e}nyi entropy,''
  [\href{\arxiv/arXiv:1803.02195}{1803.02195} [physics.gen-ph]].
  
   
  

   
    
  
  %\cite{Hsu:2004ri}
\bibitem{Hsu:2004ri} 
  S.~D.~H.~Hsu,
  %``Entropy bounds and dark energy,''
     \href{\doi/10.1016/j.physletb.2004.05.020}{Phys.\ Lett.\ B {\bf 594}, 13 (2004)}
  [\href{\arxiv/hep-th/0403052}{hep-th/0403052}].
  
    
  
  
  %\cite{Ade:2015xua}
\bibitem{Ade:2015xua} 
  P.~A.~R.~Ade {\it et al.} [Planck Collaboration],
  %``Planck 2015 results. XIII. Cosmological parameters,''
  \href{\doi/10.1051/0004-6361/201525830}{Astron.\ Astrophys.\  {\bf 594}, 
A13 (2016)}
  [\href{\arxiv/arXiv:1502.01589}{1502.01589} [astro-ph.CO]].
  
  
   
%\cite{ForemanMackey:2012ig}
\bibitem{ForemanMackey:2012ig} 
  D.~Foreman-Mackey, D.~W.~Hogg, D.~Lang and J.~Goodman,
  %``emcee: The MCMC Hammer,''
    \href{\doi/10.1086/670067}{Publ.\ Astron.\ Soc.\ Pac.\  {\bf 125}, 306 (2013)}
  [\href{\arxiv/arXiv:1202.3665}{1202.3665} [astro-ph.IM]].
  

 
   
  
%\cite{Gelman:1992zz}
\bibitem{Gelman:1992zz} 
  A.~Gelman and D.~B.~Rubin,
  %``Inference from Iterative Simulation Using Multiple Sequences,''
      \href{\doi/10.1214/ss/1177011136}{  Statist.\ Sci.\  {\bf 7}, 457 (1992)}.
 
   
  %\cite{Scolnic:2017caz}
\bibitem{Scolnic:2017caz} 
  D.~M.~Scolnic {\it et al.},
  %``The Complete Light-curve Sample of Spectroscopically Confirmed SNe Ia from 
%Pan-STARRS1 and Cosmological Constraints from the Combined Pantheon Sample,''
    \href{\doi/10.3847/1538-4357/aab9bb}{Astrophys.\ J.\  {\bf 859}, no. 2, 101 (2018)}
  [\href{\arxiv/arXiv:1710.00845}{1710.00845} [astro-ph.CO]].

  
      
%          %\cite{Suzuki:2011hu}
% \bibitem{Suzuki:2011hu}
%   N.~Suzuki {\it et al.},
%   %``The Hubble Space Telescope Cluster Supernova Survey: V. Improving the Dark Energy
% %Constraints Above z$>$1 and Building an Early-Type-Hosted Supernova Sample,''
%   \href{\doi/10.1088/0004-637X/746/1/85}{Astrophys.\ J.\  {\bf 746}, 85 (2012)}
%   [\href{\arxiv/arXiv:1105.3470}{1105.3470} [astro-ph.CO]].
  
  
  
   
%\cite{Jimenez:2001gg}
\bibitem{Jimenez:2001gg} 
  R.~Jimenez and A.~Loeb,
  %``Constraining cosmological parameters based on relative galaxy ages,''
    \href{\doi/10.1086/34054}{Astrophys.\ J.\  {\bf 573}, 37 (2002)}
      [\href{\arxiv/astro-ph/0106145}{astro-ph/0106145}].
  
  
 
   

  

\bibitem{Moresco11}
  M.~Moresco {\it et al.},
  %{\it{A 6\% measurement of the Hubble parameter at $z\sim 0.45$: direct evidence of the 
 %epoch of 
%cosmic re-acceleration}}
  \href{\doi/10.1088/1475-7516/2016/05/014}{JCAP {\bf 1605}, 014 (2016)}
  [\href{\arxiv/arXiv:1601.01701}{1601.01701} [astro-ph.CO]].
  
  
  
   
%\cite{Yu:2017iju}
\bibitem{Yu:2017iju} 
  H.~Yu, B.~Ratra and F.~Y.~Wang,
  %``Hubble Parameter and Baryon Acoustic Oscillation Measurement Constraints on the 
%Hubble Constant, the Deviation from the Spatially Flat ΛCDM Model, the 
%Deceleration–Acceleration Transition Redshift, and Spatial Curvature,''
    \href{\doi/10.3847/1538-4357/aab0a2}{Astrophys.\ J.\  {\bf 856}, no. 1, 3 (2018)}
  [\href{\arxiv/arXiv:1711.03437}{1711.03437} [astro-ph.CO]].
  
  
  %\cite{Nunes:2016qyp}
\bibitem{Nunes:2016qyp} 
  R.~C.~Nunes, S.~Pan and E.~N.~Saridakis,
  %``New observational constraints on f(T) gravity from cosmic chronometers,''
      \href{\doi/10.1088/1475-7516/2016/08/011}{JCAP {\bf 1608}, no. 08, 011 (2016)}
  [\href{\arxiv/arXiv:1606.04359}{1606.04359} [gr-qc]].
  
  
  
 
   
%\cite{Basilakos:2018arq}
\bibitem{Basilakos:2018arq} 
  S.~Basilakos, S.~Nesseris, F.~K.~Anagnostopoulos and E.~N.~Saridakis,
  %``Updated constraints on $f(T)$ models using direct and indirect measurements of the 
%Hubble parameter,''
    \href{\doi/10.1088/1475-7516/2018/08/008}{JCAP {\bf 1808}, no. 08, 008 (2018)}
  [\href{\arxiv/arXiv:1803.09278}{1803.09278} [astro-ph.CO]].
  
  
  
  
  
   
  
  
   
  \bibitem{Akaike1974} 
H. Akaike,
%{\it{A new look at the statistical model identification}}, 
\href{https://ieeexplore.ieee.org/
document/1100705/}{\emph{IEEE Transactions 
on Automatic Control}, 
\textbf{19}, (1974)  716}.


 \bibitem{bic}
 G. Schwarz, Annals of Statistics,
\textbf{6}, 461 (1978).

  
  
  
\end{thebibliography}
\end{document}